\documentclass
[aps,prl,twocolumn,floatfix,english,showpacs,10pt,superscriptaddress]{revtex4-2}%
\usepackage{graphicx}
\usepackage{amsmath}
\usepackage{physics}
\usepackage{amssymb}
\usepackage{colordvi}
\usepackage{verbatim}
\usepackage{xcolor}
\usepackage{mathrsfs}
\usepackage{epsfig}
\usepackage{lipsum}
\usepackage{amsfonts}
\usepackage[unicode=true, breaklinks=false, pdfborder={0 0 1}, backref=false,
colorlinks=true, linkcolor=blue, urlcolor=blue, citecolor=blue]{hyperref}%
\providecommand{\U}[1]{\protect\rule{.1in}{.1in}}

\setcitestyle{numbers,square}

\begin{document}   

\title{Giant Microwave Sensitivity of Magnetic Array by Long-Range Chiral Interaction Driven Skin Effect} 
\author{Tao Yu}
\email{taoyuphy@hust.edu.cn}
\affiliation{School of Physics, Huazhong University of Science and
	Technology, Wuhan 430074, China}

\author{Bowen Zeng}
\affiliation{Department of Applied Physics,
School of Physics and Electronics,
Hunan University, Changsha 410082, China}

 \begin{abstract}
 Non-Hermitian skin effect was observed in one-dimensional systems with short-range chiral interaction. Long-range chiral interaction mediated by traveling waves also favors the accumulation of energy, but has not yet showed non-Hermitian topology. Here we find that the strong interference brought by the wave propagation is detrimental for accumulation. By suppression of interference via the damping of traveling waves, we predict the non-Hermitian skin effect of magnetic excitation in a periodic array of magnetic nanowires that are coupled chirally via spin waves of thin magnetic films.  The local excitation of a wire at one edge by weak microwaves of magnitude $\sim \mu{\rm T}$ leads to a considerable spin-wave amplitude at the other edge, i.e. a  remarkable functionality useful for sensitive, non-local, and non-reciprocal detection of microwaves.

 \end{abstract}
\date{\today}

\maketitle

\textit{Introduction}.---Chiral interaction, also known as nonreciprocal coupling, refers to the asymmetric coupling amplitude between the left and right objects \cite{chiral_optics}, such as the asymmetric hopping amplitude between two nearest sites in Hatano-Nelson model \cite{Hatano_Nelson}. It has been successfully implemented to realize the non-Hermitian skin effect in one-dimensional systems, featured by a macroscopic number of eigenstates piling up at one end \cite{RMP,Zhong_Wang,Zhesen_Yang,Gong,NP_circuit,light_funneling,arbitrary_windings}. These states turn out to be topologically exceptional as showing anomalous bulk-boundary correspondence that may be characterised by generalized 
Brillouin zone \cite{Zhong_Wang,Zhesen_Yang}, and are promising for applications such as topological funneling of light \cite{light_funneling}. 

Chirality is a common ingredient in topological magnetic orders \cite{chiral_spintronics_1,chiral_spintronics_2,skyrmion_PR}, but in terms of which realization of non-Hermitian topology  is rarely addressed \cite{non_Hermitian_magnet_1,non_Hermitian_magnet_2}. Chiral interaction between  Kittel magnon of a magnetic wire (or sphere) has been recently discovered when they couple with the traveling modes such as the spin waves in films \cite{Au_simulation,chiral_spin_wave_PRB,Nano_research,magnon_trap}, waveguide microwaves \cite{chiral_waveguide_PRB,Xufeng_Zhang}, and surface acoustic waves \cite{phonon_diode,Science_Advances,Kei_phonon}, to name a few, in that the Kittel modes prefer to couple with the traveling waves propagating in one direction. We have argued that these traveling waves can mediate a long-range chiral interaction between two magnetic wires if their damping is not large \cite{magnon_trap,Nano_research}. It could be thereby speculated that these long-range chiral interaction might lead to a similar non-Hermitian skin effect to that by the chiral short-range interaction in Hatano-Nelson model \cite{Hatano_Nelson} since the energy tends to accumulate at one end. However, theoretical effort showed that these systems do not favor the coalesce of bulk modes \cite{chiral_waveguide_PRB} but only hold weak skin tendency for those modes with large decay rates, nor were the edge modes ever observed by experiments \cite{Nano_research,additional_damping,Chuanpu_wires,Grundler_wires}. The strong interference brought by the propagation of the traveling waves may be detrimental for the accumulation.

\begin{figure}[bt]
	\begin{centering}
	\includegraphics[width=8.6cm]{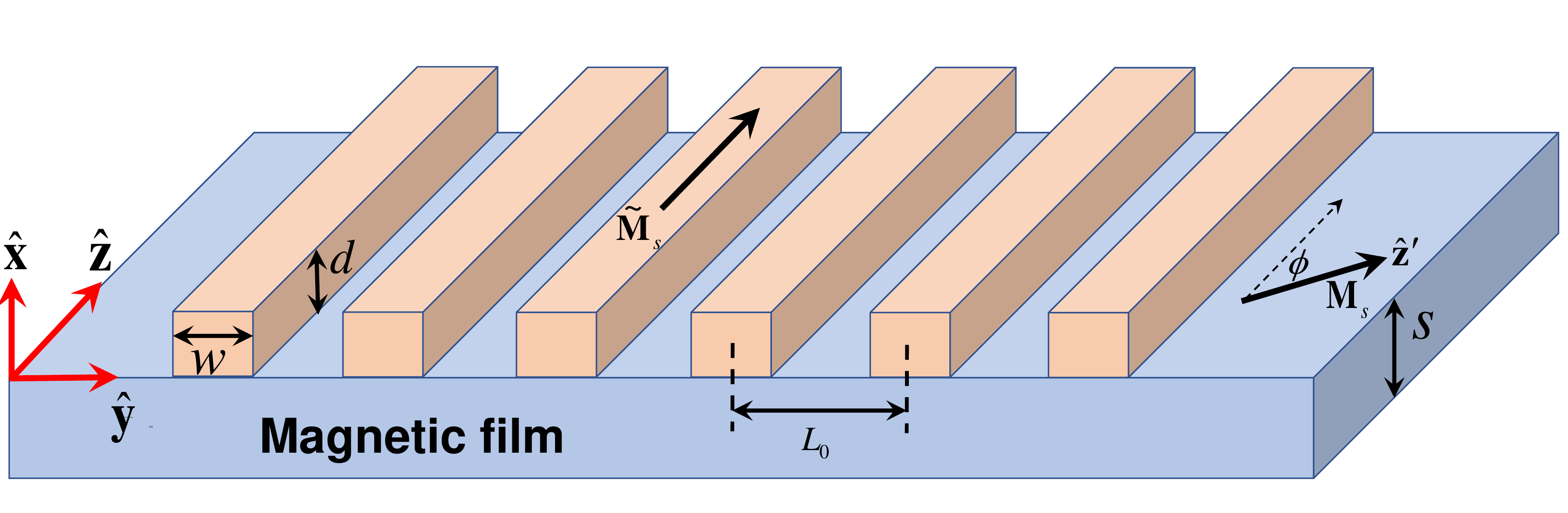}
	\par\end{centering}
	\caption{A periodic array of magnetic nanowires on top of a thin magnetic film. The direction of the saturated magnetization of the wire is pinned along the $\hat{\bf z}$ direction, while the saturated magnetization of the film is tunable by the applied magnetic field in the film plane. The geometric parameters are given in the text.}
	\label{model}
	\end{figure}

In this Letter, we propose the realization of non-Hermitian topological phase in the one-dimensional long-range coupled magnets by figuring out the collective role of the chirality and suppression of propagation interference via the damping of the interaction mediator. To be specific, we model an array of magnetic wires saturated along the wire direction on top of a thin magnetic film that are coupled via the dipolar interaction, as depicted in Fig.~\ref{model}.
When the film magnetization is along the wire direction, the coupling between the Kittel mode and the film spin waves is chiral in that the former only couples to the latter propagating in one direction  \cite{chiral_spin_wave_PRB}, the chirality being tunable by the direction of the magnetization of the film \cite{chiral_pumping}. The spin waves in the film then mediate a chiral interaction with an asymmetric coupling strength between the left and right wires. We find that when the damping of the spin waves of the film is sufficiently strong (while assuming the wire has a small damping), all the collective modes of the array of wires are localized at one edge, showing a non-Hermitian skin effect. This skin effect, however, vanishes when the damping tends to zero or the chirality is absent.  We analytically approach the generalized Brillouin zone that  characterizes a non-trivial winding of the eigen-frequency only when there exist both chirality and strong damping of the film spin waves. The non-Hermitian skin effect can act as a non-local and non-reciprocal information processor since the excitation of wire at one edge leads to a large amplitude at the other edge. It is extremely sensitive that allows for the detection of microwaves as small as $\mu{\rm T}$, a functionality that may be implemented in classical information processing and future quantum technology.

	
\textit{Chiral interaction between objects}.---We consider an one-dimensional model with a periodic array of $N$ magnetic nanowires of thickness $d$ on a thin magnetic film of thickness $s$ (Fig.~\ref{model}) \cite{Nano_research,additional_damping,Chuanpu_wires,Grundler_wires}. The distance between the neighbouring wires $L_0$ is much larger than the wire width $w$ such that the direct dipolar interaction between wires is negligible. The $l$-th wire is centered at ${\bf r}_l=R_l\hat{\bf y}=lL_0\hat{\bf y}$. The saturated magnetization $\tilde{\bf M}_s$ of the wire is pinned along the wire $\hat{\bf z}$ direction by shape anisotropy, while the film saturated magnetization ${\bf M}_s$ along $\hat{\bf z}'$ is tunable by the applied magnetic field ${\bf H}_{\rm app}$ with an angle $\phi$ with respect to the wire direction.

The interlayer exchange interaction between the wire and film is suppressed by an insulating spacer \cite{Nano_research,additional_damping}.
The magnetization $\tilde{\bf M}_l$ in the $l$-th magnetic wire couples with the stray field ${\bf h}$ from ${\bf M}$ in the film via the dipolar interaction 
$\hat{H}_{\rm int}=-\mu_0\int_0^d dxd\pmb{\bf \rho}\tilde{M}_{l,\alpha}(x,\pmb{\rho})h_{\alpha}(x,\pmb{\rho})$,
in the summation convention over repeated Cartesian
indices $\alpha=\{x,y,z\}$, $\mu_0$ being the vacuum permeability.
We disregard the nonlinear interaction between magnons when focusing on the linear regime. 
The magnetizations in the magnetic wires and film are then expanded by the magnon operator \cite{expansion_1,expansion_2},  
\begin{align}
\nonumber
&\hat{M}_{x}(\mathbf{r})=\sqrt{2M_s\gamma\hbar}\sum
_{k}\left(m_{x}^{(k)}(x)e^{iky}\hat{a}_{k}
+{\rm H.c.}\right),\\
&\hat{M}_{y}(\mathbf{r})=\cos\phi\sqrt{2M_s\gamma\hbar}\sum
_{k}\left(im_{x}^{(k)}(x)e^{iky}\hat{a}_{k}
+{\rm H.c.}\right),\nonumber\\
&\hat{\tilde{M}}_{\alpha=\{x,y\},l}(\mathbf{r})=\sqrt{2\tilde{M}_{s}\gamma\hbar}
\left(\tilde{m}_{l,\alpha}^{\rm K}({\bf r})\hat{b}_{l}
+{\rm H.c.}\right),
\label{expansion}
\end{align} 
where $\gamma$ is the
modulus of the gyromagnetic ratio, $m^{(k)}_{x}(x)$ and $\tilde{m}_{l,\alpha}^{\rm K}({\bf r})$ represent the amplitude of the spin-waves and Kittel modes, and $\hat{a}_{k}$ and $\hat{b}_{l}$ denote the  magnon operators in the film and wire. For simplicity, $k$ denotes $k_y$.
By the dipolar field of magnetic charge \cite{Jackson,chiral_pumping}, 
 the total Hamiltonian  
\begin{align}
\hat{H}/\hbar&=\sum_l\omega_{{\rm K}}\hat{b}_{l}^{\dagger}\hat{b}_{l}+\sum_{k}\omega_{k}\hat{a}^{\dagger}_{k}\hat{a}_{k}\nonumber\\
&+\sum_l\sum_{k}\left(g_{k}e^{-ikR_l}\hat{b}_{l}\hat{a}^{\dagger}_{k}+g_{k}e^{ikR_l}\hat{b}_{l}^{\dagger}\hat{a}_{k}\right)
\label{total_Hamiltonian}
\end{align}
is expressed by the coupled harmonic
oscillators.
Here, $\omega_{\rm K}$ is the frequency of the Kittel mode of the wires, $\omega_k=\mu_0\gamma H_{\rm app}+\alpha_{\rm ex}\mu_0\gamma M_sk^2$ is the dispersion of the spin waves of the film with slope governed by the exchange stiffness $\alpha_{\rm ex}$. The coupling constant
\[
g_{k}={\cal D}(k)m_x^{(k)*}\left(|k|+k\cos\phi\right)\left(\tilde{m}_{x}^{\rm K}+i{\rm sgn}(k)\tilde{m}^{\rm K}_{y}\right)
\]
depends on the propagation direction of the spin waves, the relative direction of the magnetizations in the film and nanowire, and the geometry of the wire and film that is characterized by the form factor 
${\cal D}(k)=-2\mu_0\gamma\sqrt{{M_s\tilde{M}_s}/{\Lambda}}
(1-e^{-|k|d})(1-e^{-|k|s})\sin\left({kw}/{2}\right)/{k^3}$.
Here $\Lambda$ is the length of the magnetic wire. The spin waves in the film are circularly polarized when their wavelength is sufficiently short \cite{chiral_pumping,Nano_research,additional_damping}.  Thereby when $\phi=0$ ($\phi=\pi$), i.e. the magnetization of the wire and film is parallel (anti-parallel), the wire Kittel mode only couples with the spin waves of right-going (left-going) with $g_{-|k|}=0$ ($g_{|k|}=0$)  \cite{chiral_pumping}. 

These directional spin waves can mediate a chiral interaction between two wires, approached by the Langevin equation under Hamiltonian Eq.~(\ref{total_Hamiltonian}). When the magnetic quality of the wire is higher than that of the film, we are allowed to use the Markov
approximation  \cite{input_output1,input_output2} when integrating out the film degree of freedom, yielding the Langevin equation for wires
\begin{align}
\frac{d\hat{b}_{l}}{dt}  =-i\omega_{\mathrm{K}}\hat{b}_{l}-\frac{\kappa}{2}\hat{b}%
_{l}-G_{l}(\omega)\hat{b}_{l}-\sum_{l\ne l^{\prime}}G_{ll^{\prime}}(\omega)\hat{b}_{l^{\prime}.
\label{Langevin}
}
\end{align}
It describes an effective interaction between the Kittel magnons at any instant by several coupling parameters. Here, $\kappa=2\tilde{\alpha}_{\mathrm{G}}\omega_{\mathrm{K}}$ and $\kappa_{k}=2\alpha_{\mathrm{G}}\omega_{k}$ are the Gilbert
damping of the wire Kittel modes and film spin waves, respectively, parameterized by the Gilbert coefficient $\tilde{\alpha}
_{\mathrm{G}}$ and $\alpha_{\mathrm{G}}$.
Additional damping is induced by pumping the spin waves that loses energy with rates
\[
\nonumber
G_{l}(\omega)=\sum_{k}\frac{i|g_{k}|^{2}}{\omega-\omega_{k}+i\kappa
_{k}/2}\rightarrow\frac{1}{2v(k_{\omega})}\left(  |g_{k_{\omega}}|^{2}+|g_{-k_{\omega}%
}|^{2}\right).
\]
$v(k)=2\alpha_{\rm ex}\mu_0 \gamma M_s k$ is the group velocity of the traveling waves and $k_{\omega}=\sqrt{(\omega-\mu_0\gamma H_{\rm app})/(\alpha_{\rm ex}\mu_0\gamma M_s)}$ is
the positive root of $\omega_{k}=\omega$. The spin waves mediate an effective interaction
\[
G_{ll^{\prime}}(\omega)  =i\sum_{k}e^{ik(R_{l}-R_{l^{\prime}})}%
\frac{|g_{k}|^{2}}{\omega-\omega_{k}+i\kappa_{k}/2}.
\]
We need to take into account of the finite damping of the spin waves in order to obtain the range of the interaction. To this end, we find the root of $\omega-\omega_{k}+i\kappa_{k}/2=0$ to be $q_{\omega}=k_{\omega}(1+i\alpha_G/2)$, in terms of which 
\[
G_{ll^{\prime}}(\omega)  =\frac{\Lambda}{v(k_{\omega})}e^{iq_{\omega}|l-l^{\prime}
|L_0}\left\{
\begin{matrix}
|g_{k_{\omega}}|^{2},\quad~~~R_{l}>R_{l^{\prime}}\\
|g_{-k_{\omega}}|^{2},\quad~~R_{l}<R_{l^{\prime}}%
\end{matrix}
\right.  .
\]
The interaction is of long-range when $\alpha_Gk_{\omega}L_0/2\ll 1$. The constant  $\Gamma_R=|g_{k_{\omega}}|^{2}/v(k_{\omega})$ [$\Gamma_L=|g_{-k_{\omega}}|^{2}/v(k_{\omega})$] represents the coupling strength from the left to right (right to left) wires.

\begin{figure}[th]
	\begin{centering}
	\includegraphics[width=5.6cm,trim=0cm 0.1cm 0cm 0.9cm]{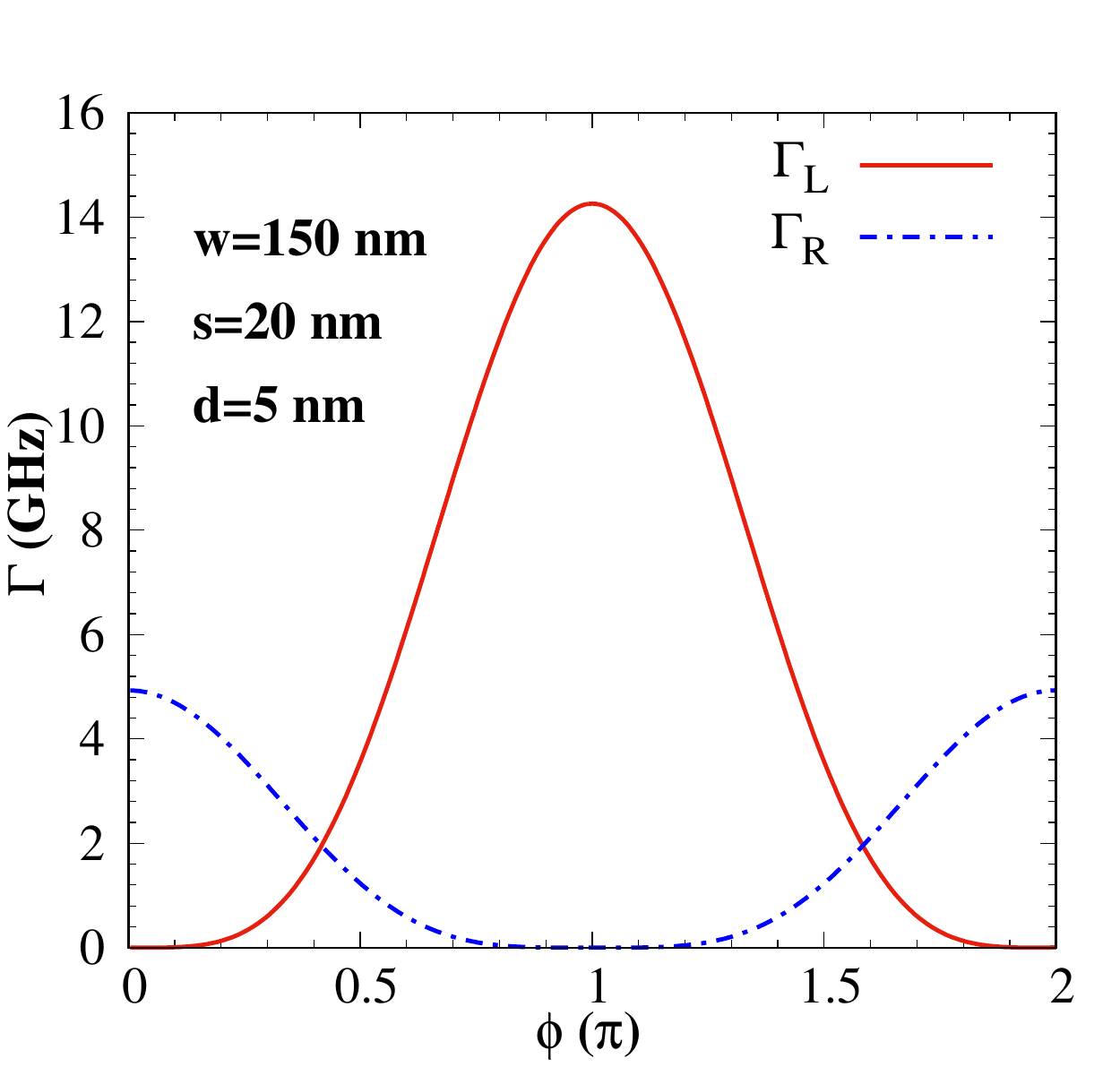}
	\par\end{centering}
	\caption{Dependence of the coupling constants $\Gamma_{L,R}$ on the direction $\phi$ of the applied magnetic field. The geometric parameters are addressed in the figure, and the material parameters are given in the text.}
	\label{fig_coupling}
	\end{figure}

We illustrate the effective couplings by exemplifying CoFeB wires of width $w=150$~nm and thickness $d=20$~nm on top of a Ni film of thickness $s=5$~nm. With $\mu_0\tilde{M}_s=0.6$~T for Ni \cite{Ni_magnetization} and $\mu_0M_s=1.6$~T \cite{CoFeB_magnetization} for CoFeB of stiffness $\alpha_{\rm ex}=8\times 10^{-13}~{\rm cm}^2$ \cite{exchange_stiffness_1,exchange_stiffness_2}, we plot the direction dependence of the coupling constants $\Gamma_{L,R}$ on the applied magnetic field of strength $\mu_0H_{\rm app}=0.1$~T in Fig.~\ref{fig_coupling}. With these parameters, the frequency of the Kittel modes of CoFeB wire is $\omega_{\rm K}=60$~GHz \cite{chiral_pumping}. 
The coupling is perfect chiral when the magnetizations of wire and film are parallel ($\Gamma_R\ne 0$ but $\Gamma_L=0$) or anti-parallel ($\Gamma_L\ne 0$ but $\Gamma_R=0$). The chirality vanishes at two critical angles $\phi_c=\{0.4\pi,1.6\pi\}$. Thereby, the system allows to simulate rich physics from with perfect to in the absence of chirality. 

\textit{Non-Hermitian skin effect}.---Conveniently, the effective non-Hermitian Hamiltonian
\begin{align}
    \hat{H}_{\rm eff}&=\left(\omega_{\rm K}-i\tilde{\alpha}_G\omega_{\rm K}-i\frac{\Gamma_R+\Gamma_L}{2}\right)\sum_{l=1}^N\hat{b}_l^{\dagger}\hat{b}_l\nonumber\\
    &-i\Gamma_R\sum_{l<l'}e^{iq_*|l-l'|L_0}\hat{b}_l^{\dagger}\hat{b}_{l'}-i\Gamma_L\sum_{l>l'}e^{iq_*|l-l'|L_0}\hat{b}_l^{\dagger}\hat{b}_{l'},
    \label{total_Hamiltonian2}
\end{align}
recovers the Langevin equation (\ref{Langevin}),
in which within on-shell approximation $\Gamma_{L,R}\equiv \Gamma_{L,R}(\omega_{\rm K})$ and $q_*\equiv q_{\omega_{\rm K}}$.
	When $\tilde{\alpha}_G<10^{-2}$ for CoFeB, the radiative damping $(\Gamma_R+\Gamma_L)/2$ by pumping the spin waves in the film dominates the damping of Kittel magnons in the wire. 
	The Hamiltonian can be expressed via a non-Hermitian matrix $\tilde{H}_{\mathrm{eff}%
	}$ via $\hat{H}_{\rm eff}=\hat{\Psi}^{\dagger}\tilde{H}_{\mathrm{eff}%
	}\Psi$, where $\hat{\Psi}=(\hat{b}_1,\hat{b}_2,...,\hat{b}_N)^T$, with matrix elements
	\begin{equation}
\tilde{H}_{\rm eff}\big|_{ll'}=
\begin{cases}
\omega_{\rm K}-i\tilde{\alpha}_G\omega_{\rm K}-i(\Gamma_{L}+\Gamma_{R})/2, & l=l'\\
-i\Gamma_{L}e^{iq_*(l-l')L_0}, & l>l'\\
-i\Gamma_{R}e^{iq_*(l'-l)L_0}, & l<l'
\end{cases}
. \label{Def:Sigma}%
\end{equation}
The phase factor in the coupling constant comes from the propagation phase of the film spin waves, thus recording the interference of waves in the range limited by $1/|q_*|$. 
Although being a generalization of Hatano-Nelson model, its topological property is, however, much less known than its short-range version. 
The right eigenvectors of $\tilde{H}_{\mathrm{eff}}$ and  $\tilde{H}_{\mathrm{eff}%
	}^{\dagger }$
	be $\{\psi _{\zeta }\}$ and $\{\phi _{\zeta }\}$ with corresponding eigenvalues $\{\nu _{\zeta }\}$ and $\{\nu
	_{\zeta }^{\ast }\}$,
	where $\zeta$ is labeled from 1 to $N$ by increasing their decay rates. $\phi _{\zeta }^{\dagger }$
	is then a left eigenvector of $\tilde{H}_{\mathrm{eff}}$. After normalization we have
	bi-orthonormality $\psi _{\zeta }^{\dagger }\phi _{\zeta ^{\prime }}=\delta
	_{\zeta \zeta ^{\prime }}$.

\begin{figure}[th]
	\begin{centering}
		\includegraphics[width=4.25cm,trim=0cm 0cm 0.1cm
1.0cm]{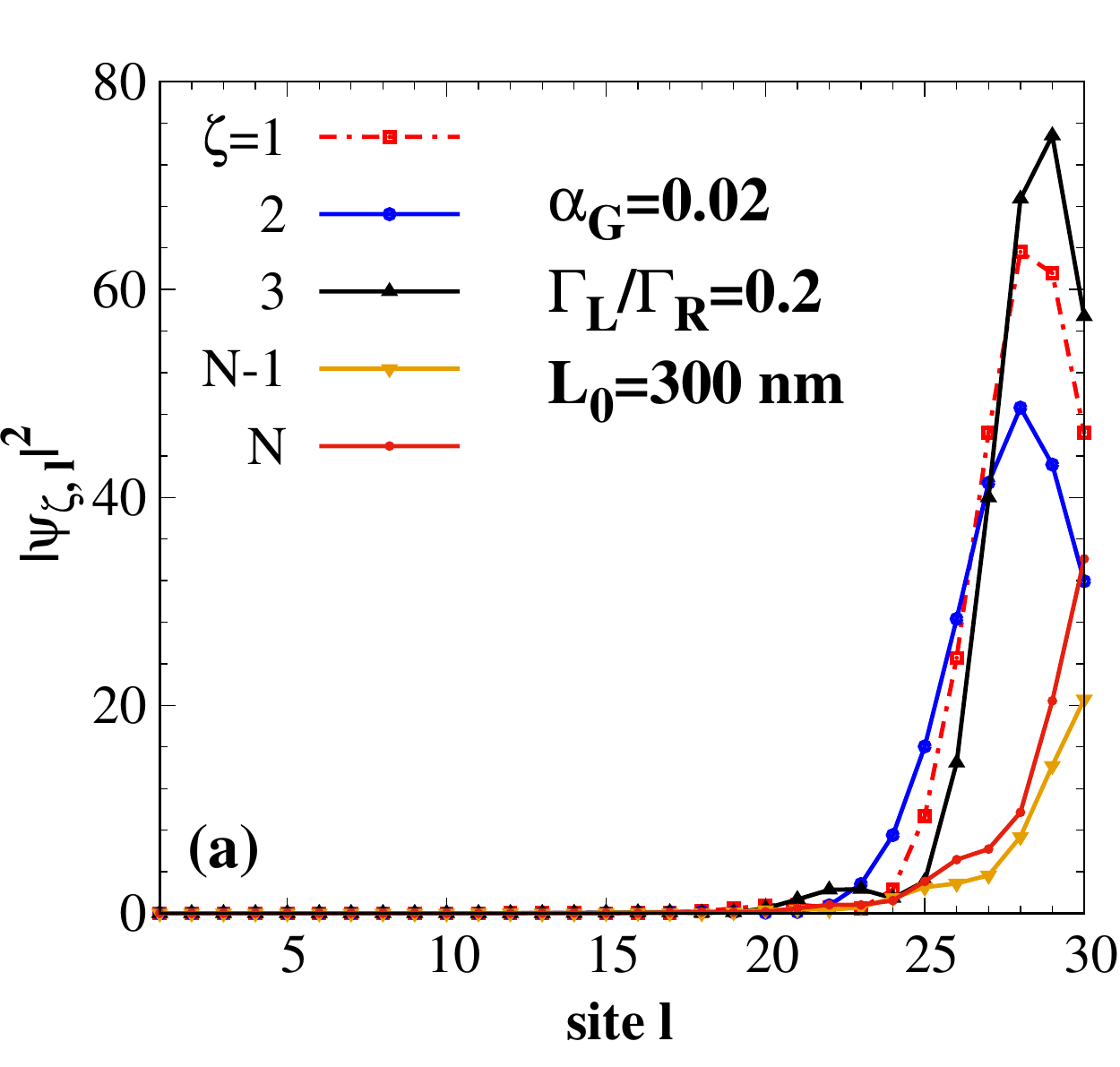}
		\includegraphics[width=4.25cm,trim=0.15cm 0cm 0cm
1.0cm]{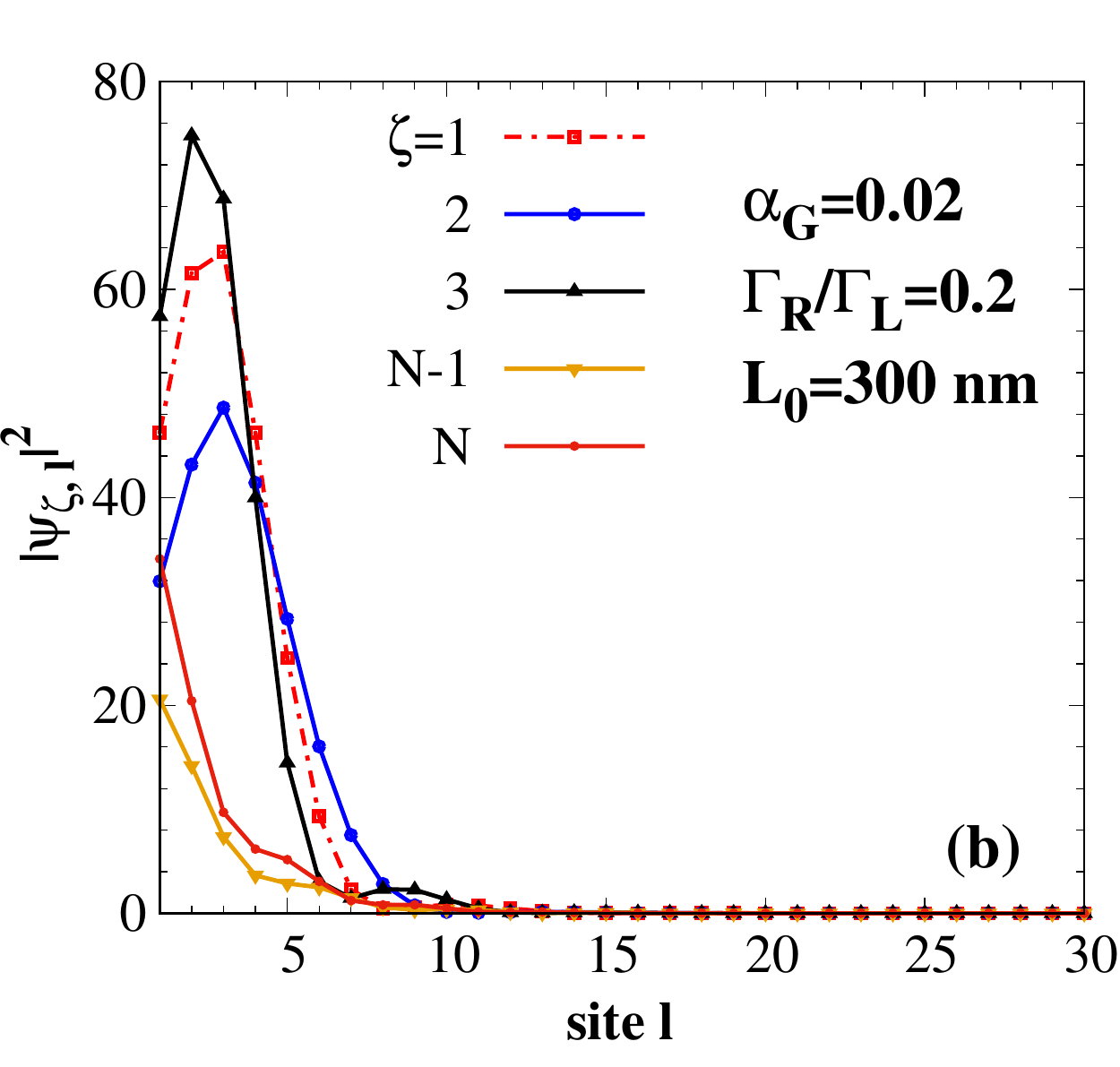}
		\includegraphics[width=4.25cm,trim=0cm 0cm 0.1cm
1.0cm]{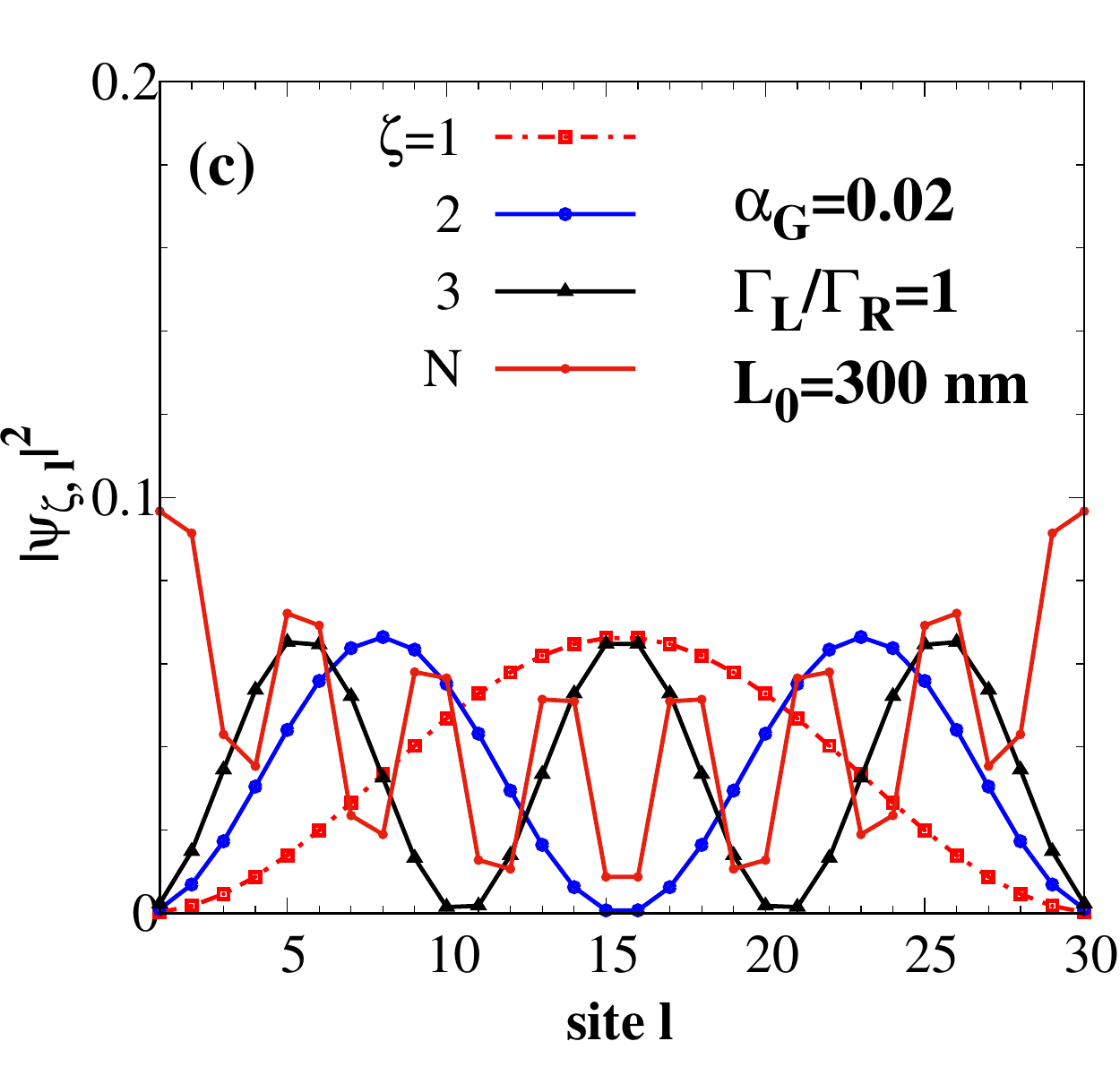}
\includegraphics[width=4.25cm,trim=0.15cm 0cm 0cm
1.0cm]{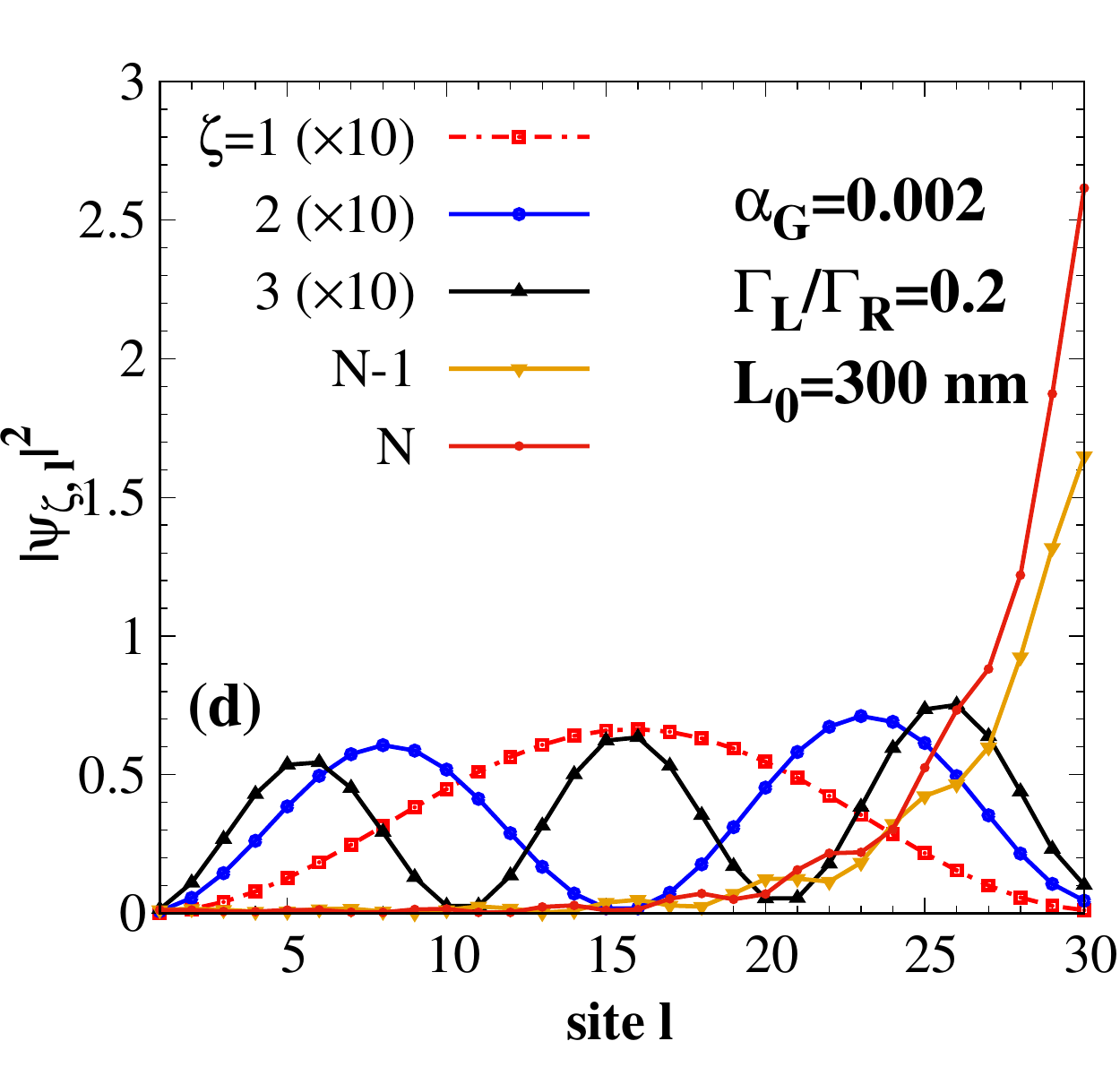}
		\par\end{centering}
	\caption{Distribution of normalized eigenmodes under different conditions. All the modes are localized at the edge in (a) and (b) when the coupling is chiral and film damping is strong. The skin modes vanish either without chirality [(c)] or with weak film damping [(d)]. }
	\label{modes}
\end{figure}

	With the material parameters in Fig.~\ref{fig_coupling}, the resonant spin waves have wave vector $k=2\pi/88.9~{\rm nm}^{-1}$. When taking the Gilbert damping $\alpha_G=0.02$ for Ni, the range of spin-wave mediated interaction is $1/{\rm Im}(q_*)=1.41~\mu{\rm m}$. The interaction is of long range by choosing the distance of neighboring wires $L_0=300$~nm. The chirality is freely tunable by changing the direction of magnetization in the film plane as in Fig.~\ref{fig_coupling}. Here we typically choose $\phi=\{0.3\pi,0.54\pi\}$ that renders $\Gamma_L/\Gamma_R=0.2$ and $\Gamma_R/\Gamma_L=0.2$ for addressing the physics.  In Fig.~\ref{modes}(a), all the modes are localized at the right edge when $\Gamma_R>\Gamma_L$, but become localized at the left edge when the chirality is reversed with $\Gamma_L>\Gamma_R$ as in Fig.~\ref{modes}(b). 
	The skin effect vanishes without the chirality at the critical angle $\phi_c=0.416\pi$, as shown in Fig.~\ref{modes}(c) that only the modes with large damping have a weak tendacy to be localized at the edge. Also, localization vanishes when taking $\alpha_G=2\times 10^{-3}$ [Fig.~\ref{modes}(d)] \cite{chiral_waveguide_PRB}. Profoundly, the mode amplitudes are enhanced by two orders in magnitude by the skin effect. This is because that these skin modes are proximity to the $N$-th order exceptional points with coalesce of all eigen-vectors \cite{RMP,exceptional_point} when one of $\Gamma_{L,R}$ is exactly zero.  

We find an analytical solution for the wavefunction that allows us to explicitly depict the generalized Brillouin zone, i.e. the distribution of complex momentum $\kappa$, parameterized by $\beta_{k}\equiv e^{ikL_0}$, on a complex plane \cite{Zhong_Wang,Zhesen_Yang}. To this end, we construct a Bloch state for a complex momentum as a traveling wave
$\hat{\Psi}_{\kappa}=({1}/{\sqrt{N}})\sum_{l=1}^N (\beta_{\kappa})^{l}\hat{b}_l$,
 obeying, under Hamiltonian (\ref{total_Hamiltonian2}), the equation of motion 
$
{d \hat{\Psi}_{\kappa}}/{dt} = -i\omega_{\kappa} \hat{\Psi}_{\kappa}-\Gamma_Lg_{\kappa}\hat{\Psi}_{q_*}+\Gamma_Rh_{\kappa}\hat{\Psi}_{-q_*}
$ \cite{Yu_xiang_1,Yu_xiang_2,chiral_waveguide_PRB}.
The dispersion relation
\begin{align}
\omega_{\kappa} &=\omega_{\rm K}-i\tilde{\alpha}_G\omega_{\rm K}-i\frac{\Gamma_R}{2}\frac{1+\beta_{\kappa}\beta_{q_*}}{1-\beta_{\kappa}\beta_{q_*}}+i\frac{\Gamma_L}{2}\frac{1+\beta_{\kappa}\beta_{-q_*}}{1-\beta_{\kappa}\beta_{-q_*}}
\label{dispersion}
\end{align}
is singular when $\kappa=\pm q_*$, implying that around these points the states have large decay rates. The traveling modes are not the eigenstates because of the existence of two edges in the chain that radiates energy with amplitudes
\begin{equation}
	g_{\kappa}=\frac{1}{1-\beta_{\kappa}\beta_{-q_*}},\ \ h_{\kappa}=\frac{(\beta_{\kappa})^N(\beta_{q_*})^{N}}{1-\beta_{\kappa}\beta_{q_*}},
\end{equation}
and reflects the traveling modes. Thus we may superpose two traveling modes of the same energy with different momenta, i.e.
\begin{align}
    \omega_{\kappa}=\omega_{\kappa'}
    \label{relation_1}
\end{align}
for a new mode.
Superposition $\hat{\Delta} = g_{\kappa'} \hat{\Psi}_{\kappa} - g_{\kappa} \hat{\Psi}_{\kappa'}$ obeys
\begin{equation}
\frac{d \hat{\Delta}}{dt}  =-i \omega_{\kappa} \hat{\Delta} + \Gamma_R\left(g_{\kappa'} h_{\kappa} - g_{\kappa} h_{\kappa'}\right)\hat{\Psi}_{-q_*},
\end{equation}
and becomes the eigenmode when
 \begin{align}
     g_{\kappa'} h_{\kappa}=g_{\kappa} h_{\kappa'}.
     \label{relation_2}
 \end{align}
Equations (\ref{relation_1}) and (\ref{relation_2}) are the desired relations to find the complex momentum $\kappa$, substituting which into Eq.~(\ref{dispersion}) leads to the dispersion. Numerically diagonalizing the Hamiltonian with eigen-frequency $\omega_{\kappa}$ solves the complex momentum  
\begin{align}
    \beta_{\kappa}^{(\pm)}=(-B_{\kappa}\pm \sqrt{B_{\kappa}^2-4A_{\kappa}C_{\kappa}})/(2A_{\kappa}),
    \label{roots}
\end{align}
where with $\tilde{\omega}_{\kappa}\equiv\omega_{\kappa}-\omega_{\rm K}+i\tilde{\alpha}_G\omega_{\rm K}$
\begin{align}
    A_{\kappa}&=\tilde{\omega}_{\kappa}-i(\Gamma_R-\Gamma_L)/2,\nonumber\\
    B_{\kappa}&=-\left(\tilde{\omega}_{\kappa}-i\frac{{\Gamma_R}+\Gamma_L}{2}\right)\beta_{q_*}-\left(\tilde{\omega}_{\kappa}+i\frac{{\Gamma_R}+\Gamma_L}{2}\right)\beta_{-q_*},\nonumber\\
    C_{\kappa}&=\tilde{\omega}_{\kappa}+i(\Gamma_R-\Gamma_L)/2.
    \nonumber
\end{align}
Equation~(\ref{roots}) contains two roots of momentum $\kappa$ and $\kappa'$ at the same frequency, confirming the relation (\ref{relation_1}).

\begin{figure}[th]
	\begin{centering}
		\includegraphics[width=4.45cm,trim=0cm 0cm 0cm
0.5cm]{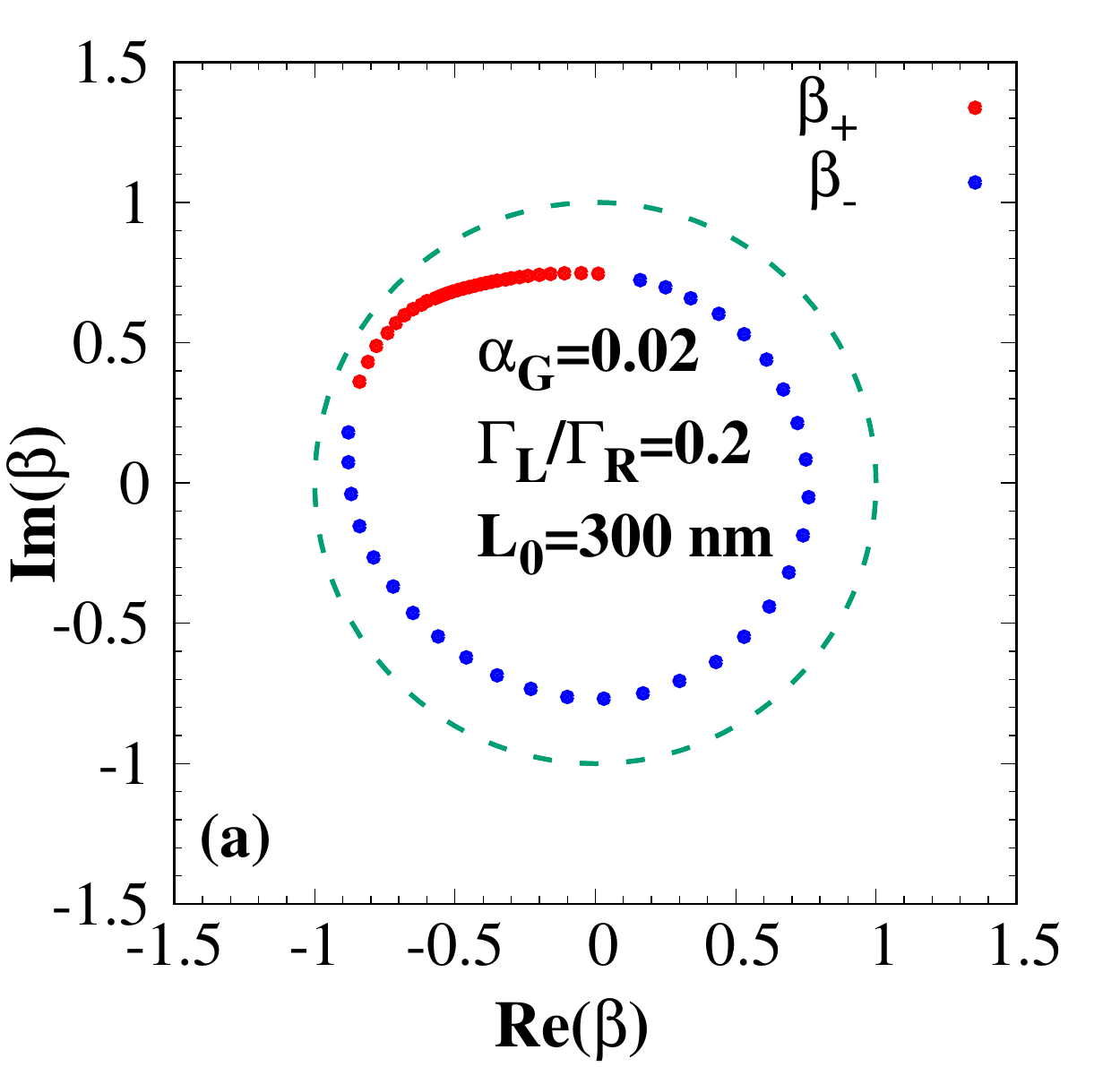}
\hspace{-0.5cm}
		\includegraphics[width=4.45cm,trim=0cm 0cm 0cm
0.5cm]{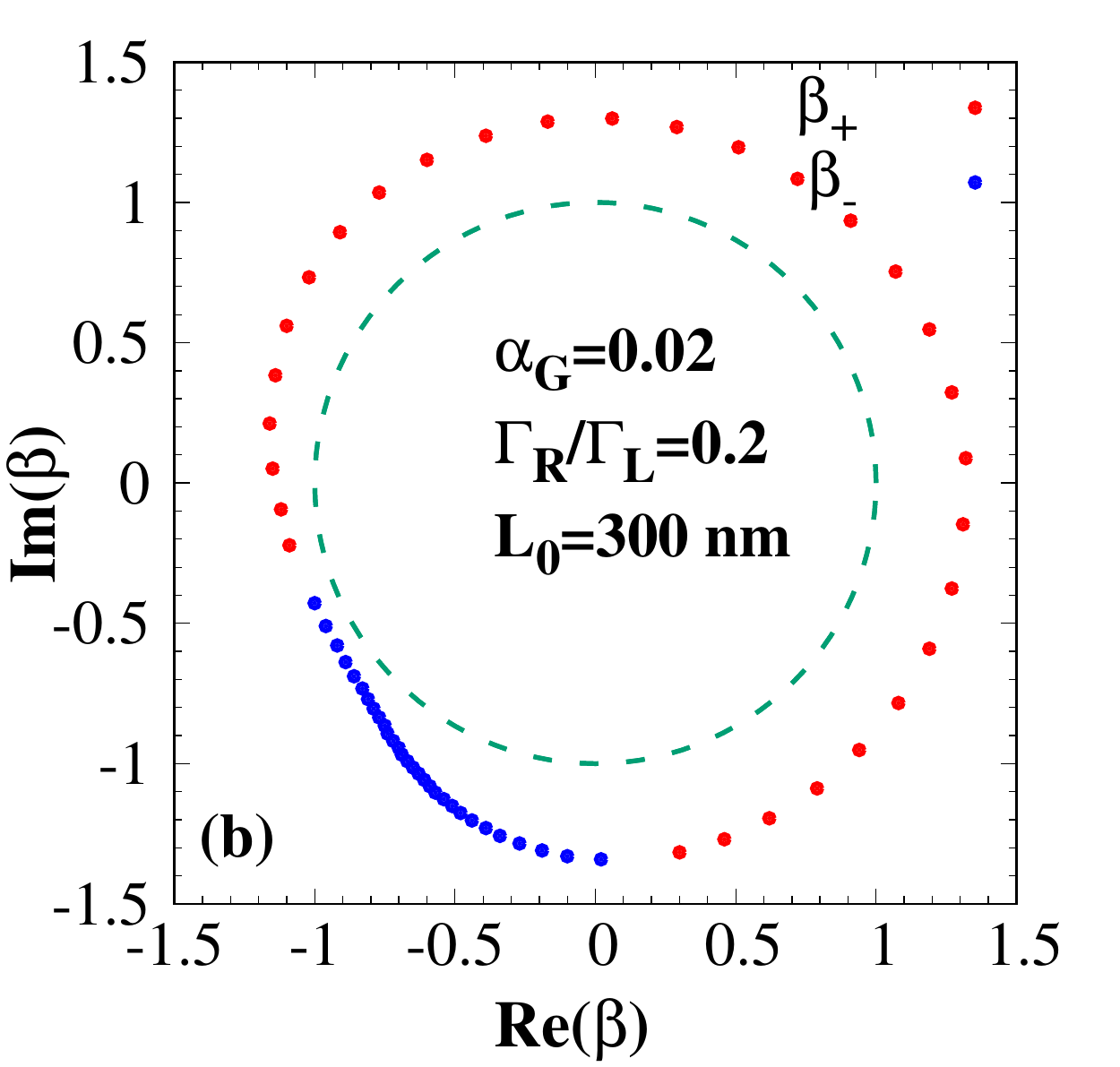}\\
		\includegraphics[width=4.45cm,trim=0cm 0cm 0cm
0.5cm]{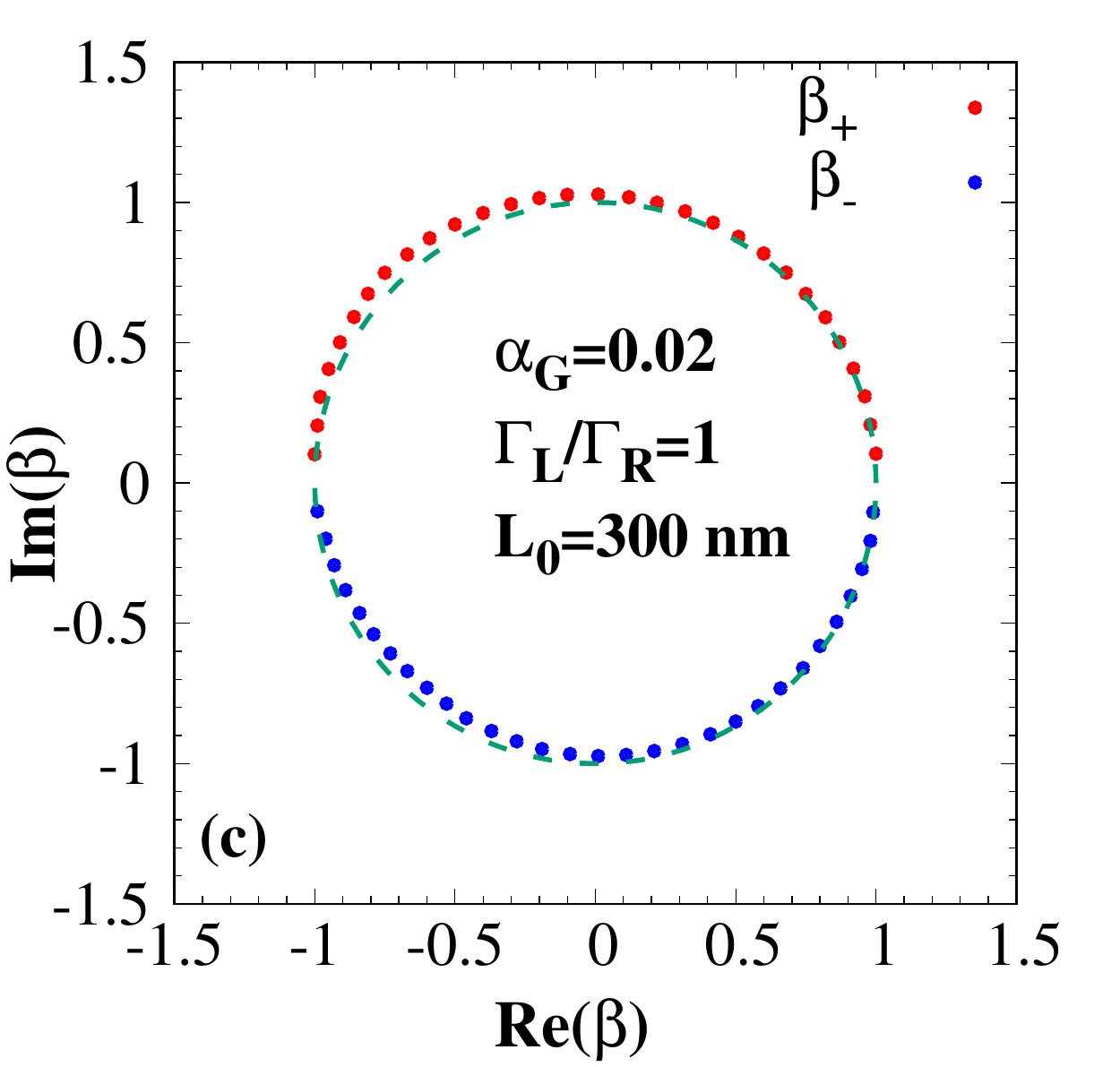}
\hspace{-0.5cm}
\includegraphics[width=4.45cm,trim=0cm 0cm 0cm
0.5cm]{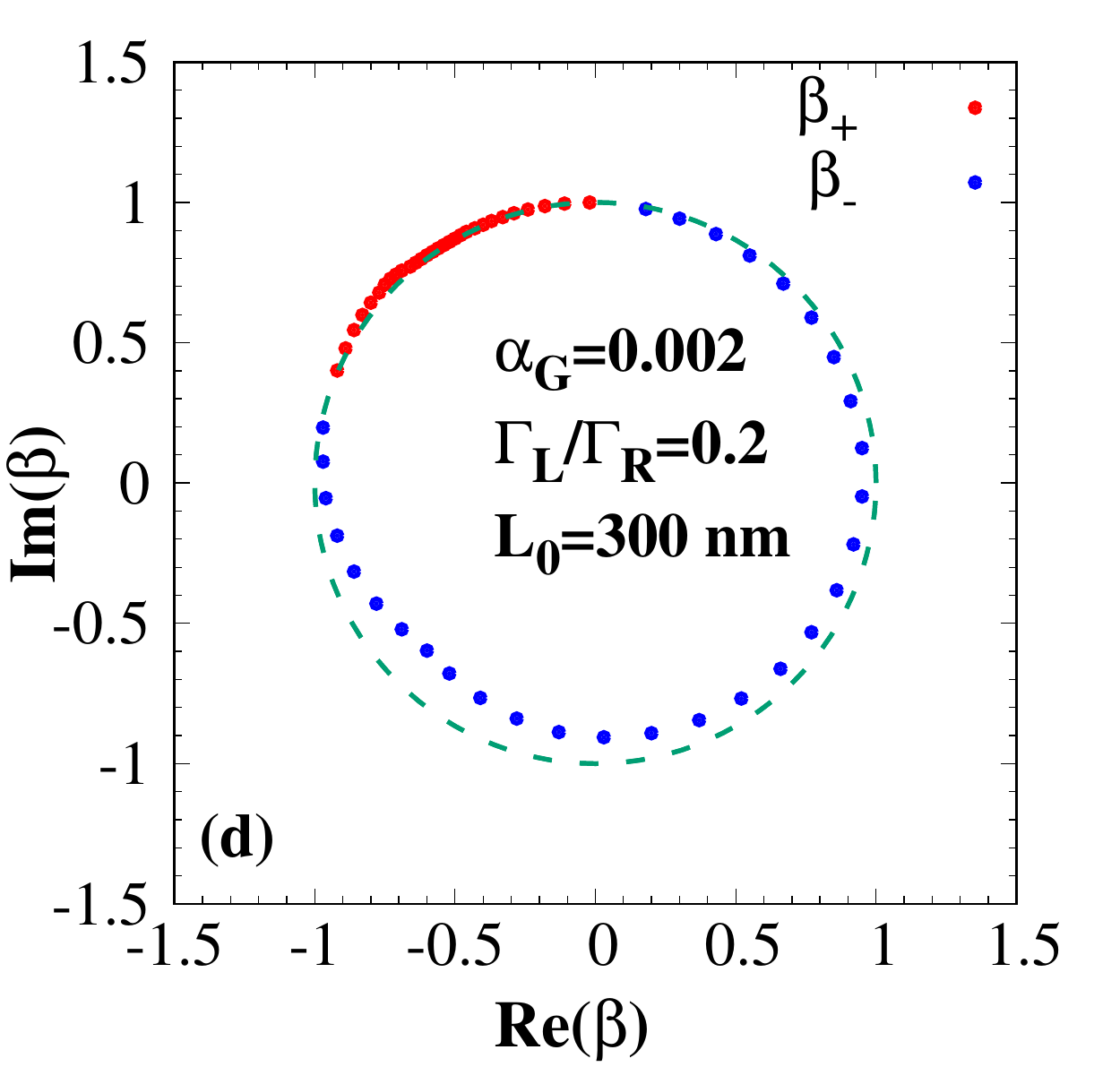}
		\par\end{centering}
	\caption{Generalized Brillouin zone under different chiralities and dampings. $|\beta|<1$ ($|\beta|>1$) in (a) and (b) favors the localization at the right and left edges, respectively, while $|\beta|\approx 1$ in (c) and (d) indicates the absence of the skin effect.}
	\label{GBZ}
\end{figure}

On the other hand, for the eigenmodes we expand $\hat{\Delta}=\sum_l\phi_{\zeta,l}^*\hat{b}_l$ and find the  wavefunction
\begin{align}
\psi_{\zeta,l}=C\left(g_{\kappa'}\beta_{\kappa}^{N-l}-g_{\kappa}\beta_{\kappa'}^{N-l}\right),    
\end{align}
to be normalized with a constant $C$. The exponent $(N-l)$ controls the distribution of excited wire magnons. When $|\beta_{\kappa}|>1$ ($|\beta_{\kappa}|<1$), the amplitude of $\psi_{\zeta,l}$ decreases (increases) with increasing the sites from $1$ to $N$, implying the localization at the left (right) edge of the chain.
Figure~\ref{GBZ} plots the distribution of the real and imaginary parts of $\beta_{\pm}$, which form a loop in the complex plane, under different conditions. When there are net chirality and strong damping of the film, $|\beta_{\pm}|$ labeled by the red and blue dots deviate strongly from unit that is indicated by the green dashed line. This is the condition for the emergence of the non-Hermitian skin effect \cite{Zhong_Wang,Zhesen_Yang}. When the chirality vanishes or the damping of film becomes small, the distribution of $\beta$ almost overlaps with the unit circle, indicating the absence of skin effect.

\begin{figure}[th]
	\begin{centering}
		\includegraphics[width=5.6cm,trim=0cm 0.1cm 0cm
0.5cm]{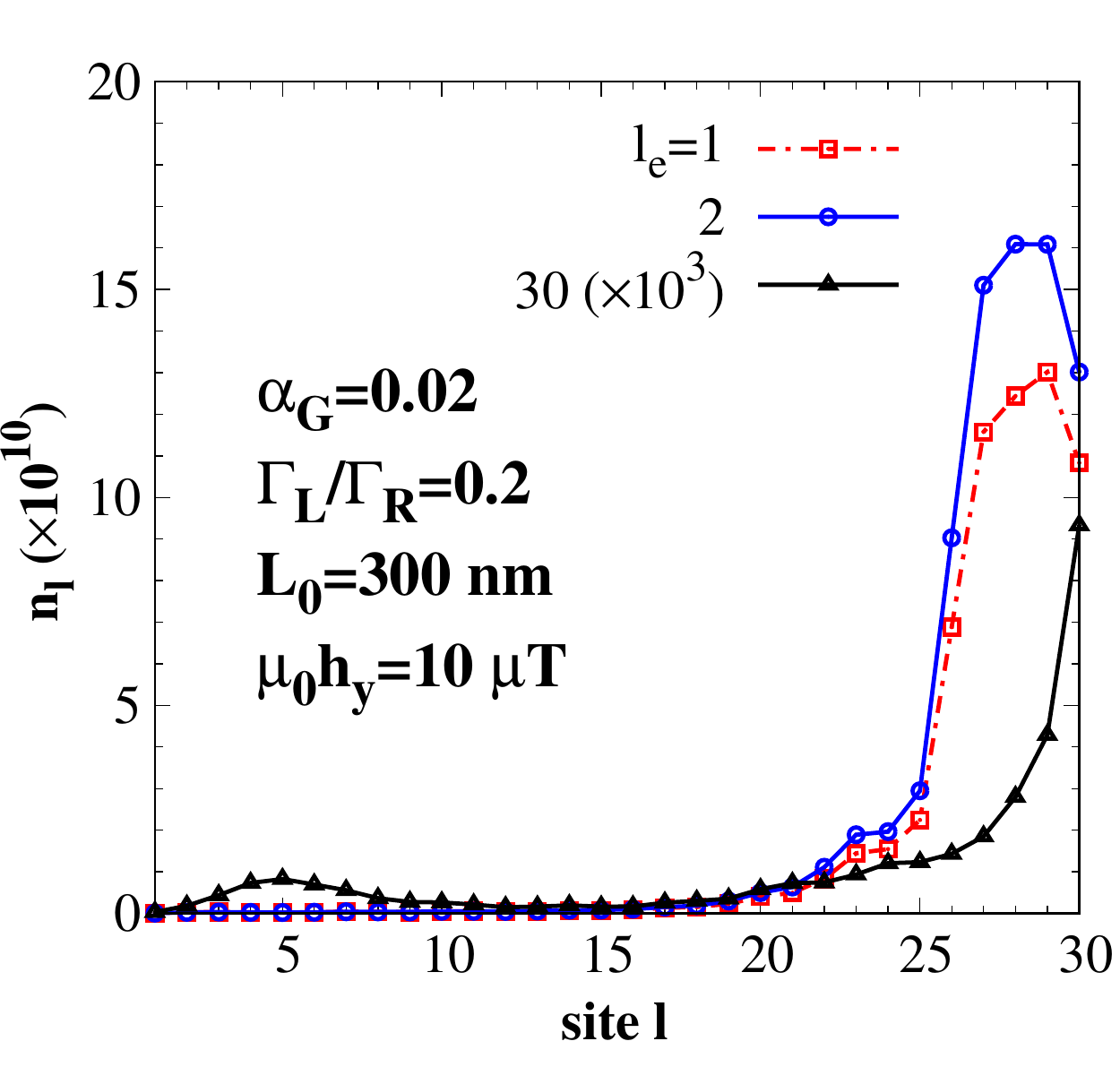}
		\par\end{centering}
	\caption{Non-local excitation of magnons by weak microwaves. The excited magnons accumulate at the right edge when the stripline locally excites the $l_e$-th wire.}
	\label{Excitation}
\end{figure}

\textit{Sensitive detection of microwaves}.---The wires can be excited and detected by the local metal stripline with comparable width to $w$ that is assumed to support an uniform electric current in the cross section \cite{Nano_research}. A thin stripline on top of the $l_e$-th wire generate a magnetic field $h_y\hat{\bf y}$ of frequency $\omega_d$ that locally excite the beneath wire. The interaction between the field and wire Kittel mode $\hat{H}_{\rm int}'=\hbar (g_{l_e}e^{-i\omega_d t}\hat{\beta}_{l_e}^{\dagger}+{\rm H.c.})$ is parameterized by
$g_{l_e}=-\mu_0\sqrt{\tilde{M}_s\gamma/(2\hbar)}h_yd w \tilde{m}_{l_e,y}^{\rm K}$.
At the steady-state, the excited amplitude of the magnetization in every wire follows 
$\langle \hat{\Psi}\rangle=\sum_{\zeta}{g_{l_e}\phi_{\zeta,l_e}^{\dagger}}\psi_{\zeta}/{(\omega_d-\gamma_{\zeta})}$.
To be realistic, we take into account of the disorder modelled by the random shift $\delta\omega\in [-0.01\omega_{\rm K},0.01\omega_{\rm K}]$ to the Kittel frequency of every wire. The skin effect turns out to be robust to this disorder. Only launching of the opposite edge can efficiently excite the localized modes at the edge, i.e. non-local and non-reciprocal excitation. On the other hand, the mode amplitudes are giantly enhanced by the skin effect as in Fig.~\ref{modes}, leading to the expectation of sensitive microwave detection. Figure~\ref{Excitation} is the numerical substantiation of the above expectation: a small microwave field $10~{\mu}{\rm T}$ of frequency $\omega_d=60$~GHz leads to a deviation of the magnetization $2\mu_0\sqrt{2\tilde{M}_s\gamma\hbar n_{l}}\tilde{m}_{l,y}=0.13$~T of CoFeB wire with $\tilde{\alpha}_G=0.002$ at the steady state, i.e. a precession cone angle of $\sim 4.7^{\circ}$, $n_l$ being the excited magnon number in the $l$-th wire. The results converge when averaging up to $10^{4}$ samples with random disorder.

\textit{Discussion}.---To conclude, we predict the non-Hermitian skin effect of chirally coupled magnetic wires that are mediated by the spin waves of significant damping, which overcomes the strong interference brought by the propagation of the traveling waves. This leads to functionalities such as non-local spin-wave excitation and giantly sensitive detection of microwaves, and makes it easier to achieve nonlinear regime of magnons with a small power. The generalization of our scenario to chiral photonics \cite{chiral_optics,nano_optics}, plasmonics
\cite{plasmonics_1,plasmonics_2,Nori}, and acoustics \cite{acoustic_1,acoustic_2} may promote the performace of sensors for detection of small signals.
	
\vskip0.25cm	
\begin{acknowledgments}
	This work is financially supported by the National Natural Science Foundation of China and the startup grant of Huazhong University of Science and Technology (Grant No. 3004012185). B. Z. is  financially supported by the National Natural Science Foundation of China Grant No. 12004106. We thank Gerrit E. W. Bauer, Yu-Xiang Zhang, and Jian-Song Pan for useful discussions. 
\end{acknowledgments}

	\end{document}